\documentclass[12pt]{iopart}

\usepackage{graphicx}
\usepackage{color}
\usepackage{bm}         
\usepackage{textcomp}
\usepackage{psfrag}
\usepackage{iopams}
\usepackage[mathscr]{eucal}

\newcommand {\bkt} [1] {\langle #1 \rangle}
\newcommand {\beq}{\begin{equation}}
\newcommand {\eeq}{\end{equation}}
\newcommand{\bea}{\begin{eqnarray}}
\newcommand{\eea}{\end{eqnarray}}
\newcommand {\pd} [2] {\frac{\partial #1}{\partial #2}}
\newcommand {\td} [2] {\frac{d #1}{d #2}}

\begin{document}

\title{Condensation of actin filaments pushing against a barrier}
\author{Kostas Tsekouras$^{1,2}$, David Lacoste$^1$, Kirone Mallick$^3$ and Jean-Fran\c{c}ois Joanny$^2$}
\address{$^1$ Laboratoire de Physico-Chimie Th\'eorique - UMR CNRS Gulliver 7083,
ESPCI, 10 rue Vauquelin, F-75231 Paris, France}
\address{$^2$ Physico-Chimie UMR 168, Institut Curie, Paris, France}
\address{$^3$ Service de Physique Th\'eorique, Commissariat \`a l'Energie Atomique- Saclay, Gif, France}

\ead{tamytes8a@gmail.com}

\begin{abstract}
We develop a model to describe the force generated by the polymerization of an array
of parallel biofilaments.
The filaments are assumed to be coupled only through mechanical contact with a movable barrier.
We calculate the filament density distribution and the force-velocity relation with a mean-field approach combined with simulations.
We identify two regimes: a non-condensed regime at low force in which filaments are spread out spatially, and a condensed regime at high force in which filaments accumulate near the barrier.
We confirm a result previously known from other related studies, namely that the stall force is equal to $N$ times the stall force of a single filament.
In the model studied here,
the approach to stalling is very slow, and the velocity is practically zero
at forces significantly lower than the stall force.
\end{abstract}

\submitto{\NJP}
\maketitle

\section{Introduction}
Actin filaments and microtubules are key components of the cytoskeleton of eukaryotic cells.
Both play an essential role for cell motility and form the core components of
various structures such as lamellipodia or filopodia. They are active elements which
exhibit a rich dynamic behavior. For instance, actin filaments treadmill in a process where
monomers are depolymerized from one end of the filament while other monomers are repolymerized at the other end. Actin polymerization is highly regulated in the cell, through many actin binding proteins. Some of these proteins accelerate actin polymerization, while others crosslink filaments or create new branches from existing filaments. All these proteins ultimately control the force that a cell is able to produce \cite{pollard09}.

Given the complexity of actin polymerization, many studies have focused on its basic structural element, namely the filament itself.
For instance, a lower bound for the polymerization force generated by a single actin filament has been deduced from the buckling of a filament which was held at one end by a formin domain and at the other end by a myosin motor \cite{kovar_pollard}. Other studies focused on the
dynamics of single filaments through depolymerization experiments \cite{mitchison}.
In order to understand the rich dynamical behavior of single filaments like actin or microtubules and the force they can generate,
discrete stochastic models have been developed which incorporate at the molecular level the coupling of hydrolysis and polymerization  \cite{kolomeisky05,kolomeisky06,david09,david10,vavylonis05,antal-ideal,antal-instab,lipowsky}.
The filament dynamics and the force generation are two related aspects: Hydrolysis is not only relevant for understanding the single filament dynamics but also for the force generation, since the force generated by a filament is typically lowered by hydrolysis \cite{david09}.

Ensembles of parallel interacting filaments are able to generate larger forces than single filaments as in cellular structures called filopodia \cite{altigan_sun}.
General thermodynamic principles controlling the force produced by the polymerization of growing filaments pushing against a movable barrier were put forward many years ago by Hill et al. \cite{hill_kir}. For many years however, it was unclear how to extend these results in order to understand theoretically the effect of interactions or collective effects in the process of force generation. Progress in this direction was made through the introduction of stochastic models for ensembles of parallel microtubules \cite{mog_ost,van-doorn,tanasethesis}, and through the development of simulations for actin filaments in parallel geometry \cite{note} or in networks \cite{carlsson}. In these works, the brownian ratchet model \cite{peskin_oster93} was used at the single filament level, while some specific rule was assumed concerning the way the load is shared between the filaments.
In the absence of hydrolysis and lateral interactions between the filaments, the stall force of an ensemble of parallel $N$ filaments should be $N$ times the stall force of a single filament, as confirmed either by a detailed balance argument valid only near stalling \cite{van-doorn} or more recently by a more general analysis based on a decomposition into cycles \cite{kierfeld}.

In this work, we propose a new theoretical framework for this problem. One novel aspect as compared to previous work \cite{van-doorn} comes from the fact that we model the dynamics of an ensemble of parallel non-interacting filaments at an arbitrary value of the force, rather than just predict the value of the stall force. Another important difference between this model and previous work, is that our model allows an arbitrary number of filaments in contact with the moving wall, which allows the possibility of a condensation transition for the number of filaments at the wall.

This paper is organized as follows: we first present the model, secondly
the mean-field approach for the general case of an arbitrary $N$, then the simulations, and a theoretical analysis of the approach to stalling. We end with a discussion of various related experiments in this field, in which forces generated by a few actin filaments have been measured \cite{footer,experiment}.

\section{Model}
We consider two rigid flat surfaces: one fixed where filaments are nucleated (nucleating wall) and one movable (barrier) whose position is defined to be the position of the filament(s) furthest away from the nucleating wall (thus there is always at least one filament in contact with the barrier).
In the cellular environment, this ``barrier" is often a membrane against which filaments exert mechanical forces.
We do not model the internal structure of the filaments, and in particular we do not account for ATP hydrolysis.
After nucleation, the filaments grow or shrink by exchanging monomers with the surrounding pool of monomers, which
acts as a reservoir. The filaments are coupled only through mechanical contact with the barrier.
In some previous models \cite{mog_ost}, a staggered distribution of initial filaments was assumed so that there would be
only a single filament in contact at a time. Here we do not make such an assumption, on the contrary monomers inside different filament are precisely lined up. As a result, the number of filaments at contact is an arbitrary strictly positive integer.

\begin{figure}
   \begin{center}
      \includegraphics[scale=0.5]{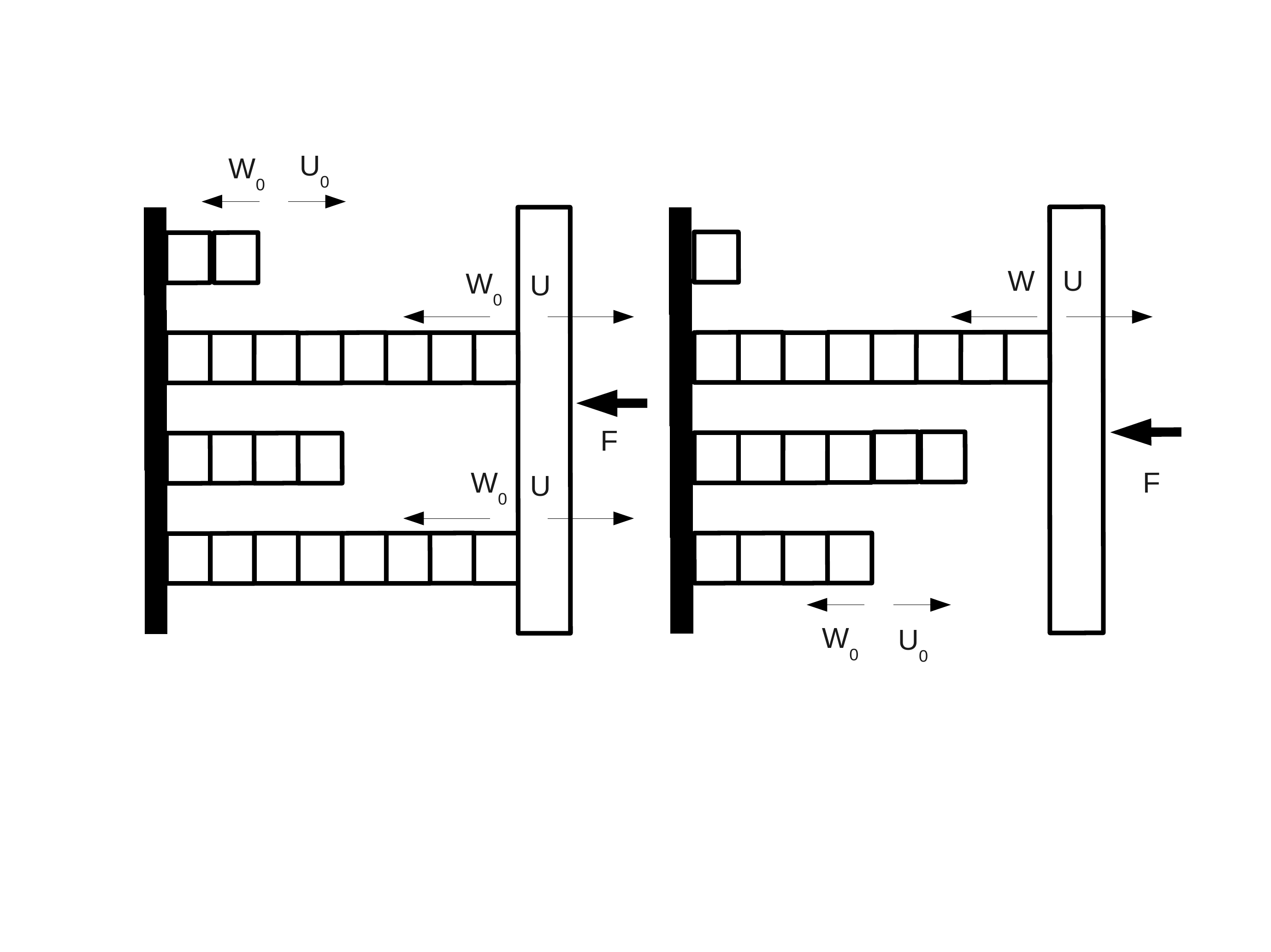}
\caption{Representation of the filaments pushing on a barrier (the white vertical rectangle on the right, which exerts a force $F$ on the filaments). The right figure corresponds to the case that only one filament is in contact with the barrier while the left figure corresponds to the case where several filaments are in contact with the barrier. The on and off rates of monomers onto free filaments are $U_0$ and $W_0$. The on-rate on filaments  in contact is $U$, and the off-rate is $W$ when there is only one filament in contact and $W_0$ otherwise.}
      \label{fig:sketch}
   \end{center}
\end{figure}

It follows that we can separate the filaments in two populations, the free filaments which are not in contact with the barrier,
and the filaments in contact. Only the filaments in contact  feel the force exerted by the barrier on them, and
as a result this changes their polymerization rates as compared with free filaments.
We assume that a monomer can be added to any free filament with rate $U_0$ or removed with rate $W_0$, as shown in Fig\ref{fig:sketch}. Similarly, a monomer can be added to a filament in contact with rate $U(F)$, and removed with a rate $W(F)$ (or $W_0$ as explained below).
The values of the rates which we have used correspond to an actin barbed end and are given in table \ref{key}. We also assume that the barrier exerts a constant force $F$ on the filaments in contact, this force is defined to be positive when the filaments are compressed.

We need now to specify more precisely how the force exerted by the barrier is shared by the filaments in contact. When a monomer is
added to a filament in contact, the barrier moves by one unit, but only the filament on which
the monomer has been added does work; we therefore treat all the other filaments as free during that step.
Similarly, during depolymerization, filaments depolymerize from the barrier with the free depolymerization
rate $W_0$ as long as there is at least one other filament in contact with the barrier, since in this case the depolymerizing
filaments do not produce work. The depolymerization occurs with a rate $W$ only when there is a single filament in contact with the barrier. In this case the filament produces work, since its depolymerization leads to the motion of the barrier.

For a filament which has exchanged work with the barrier through addition or loss of monomers, we use a form of local detailed balance which reads :
\beq \label{eqn:rateratio} \frac{U}{W}=\frac{U_0}{W_0}e^{-f}. \eeq
This relation is obeyed by the following parametrization of the rates \cite{van-doorn,hill-book,bell}:
\beq \label{eqn:rates} U=U_0e^{-f\gamma}\mbox{ and }W=W_0e^{f(1-\gamma)},\eeq
where $\gamma$ is the ``load factor'' and $f$ is the non-dimensional force $f=Fd/k_BT$, where d is
the monomer length. Note that $\gamma$ itself could be a function of the force, however in the
following we assume that it is just a constant. More elaborate treatments of the load dependence of the transition rates can be found for instance in Ref.~\cite{walcott}.

An essential feature of this model is that although multiple filaments interact with the barrier,
when a monomer is added to one of the filaments in contact, it must do work against the entire load.
In the classification of \cite{borisy}, this corresponds to a scenario with ``no load sharing".
If the force could be shared by more than one filament or if the monomers in different filaments were not precisely lined up, the above discussion would still apply: in this case a single filament would carry a fraction of the load at a time, and for that filament a similar local detailed balance would hold.
In this case, although the stall force would be the same as in the  "no-load sharing" scenario,  the form of the force-velocity curve would
be affected. Such models have been considered in Refs.~\cite{van-doorn,tanasethesis,borisy,kierfeld},
but for simplicity, in the present paper, we focus on the ``no load sharing" model.

\section{Theory}
In the particular case that there are only two filaments ($N=2$), the master equation can be solved exactly in terms of the probability that there is a given gap at a given time between the two filaments, as shown Appendix A. Unfortunately, this approach is limited to the $N=2$ case, because only in that case there is a single gap between the filaments. For $N>2$, there
are many gaps, so in general such an approach quickly becomes as complicated as the one based
on the filaments themselves. So instead of looking for an exact solution, we provide in the section below, an approximate but accurate mean-field solution for the general case $N>2$.

\subsection{An ensemble of $N$ filaments with $N>2$}
We recall that the position of the moving barrier coincides with the position of the longest
filament, and we define $N_i$ as the number of filament ends, which are present at a distance $i$ from the moving barrier.
We take the convention that $i=0$ corresponds to the barrier itself.
Since each filament has only one active end and the total number of filaments is fixed to be $N$,
we have the condition that $\sum_{i=0} N_i=N$.
After a careful account of all the possible events that can occur on any filament in a small time interval,
we obtain the following master equations:
\beq\label{eqn:mas1}\td{N_i}{t}=(W_0+UN_0)N_{i-1}+(U_0+W\delta_{N_0=1})N_{i+1}-(W_0+U_0+UN_0+W\delta_{N_0=1})N_i,\eeq
\beq\label{eqn:mas2}\td{N_1}{t}=(U_0+W\delta_{N_0=1})N_2-(U_0+W_0+W\delta_{N_0=1}+UN_0)N_1+[W_0(1-\delta_{N_0=1})+U(N_0-1)]N_0, \eeq
\beq\label{eqn:mas3}\td{N_0}{t}=(U_0+W\delta_{N_0=1})N_1-[U(N_0-1)+W_0(1-\delta_{N_0=1})]N_0, \eeq
where $\delta_{N_0=1}$ represents the probability that there is only a single filament in contact.

In deriving these equations, we have for instance implicitly replaced the joint probability to have $N_i$ filament ends at position $i$ and to have only one filament at contact at time $t$, namely $P(N_i(t)=N_i,N_0(t)=1)$ by the product of $P(N_i(t)=N_i)$ and $P(N_0(t)=1)$. In other words, a mean-field approximation has already been used. A further consequence of this mean-field approximation is that in these equations, $\delta_{N_0=1}$ can be replaced by its time-averaged value, which we call $q$:
\beq \label{self-consistent eq} q=\bkt{\delta_{N_0=1}}. \eeq
The quantity $q$ is a central feature of our model for $N>2$. All subsequent results and calculations appearing in this paper follow from this mean-field approximation.

At steady state, the l.h.s. of Eq.~\ref{eqn:mas1} is zero. The r.h.s. leads to a recursion valid for $i \ge 2$, which
can be solved after a few lines of calculations. The solution is
\beq
N_i=N_2 \exp(-(i-2)/l),
\label{eqn:profile gen case}
\eeq
where $l$ is the correlation length (expressed in number of subunits)
given by \beq \label{eqn:charL} l=\left[\ln\left(\frac{U_0+Wq}{W_0+UN_0}\right)\right]^{-1}. \eeq
The other two equations Eqs.~\ref{eqn:mas2}-\ref{eqn:mas3}, together with the normalization condition fix $N_2, N_1$ and $N_0$.
We find that the average number of filaments in contact with the wall $N_0$ is:
\beq \label{eqn:fullN0} N_0=\frac{(U_0+Wq-W_0)N}{U_0+U(N-1)+(W-W_0)q}.\eeq

When $N=2$, this mean-field solution agrees with the exact solution derived
in appendix A with the additional condition that $\gamma=1$, in which case the on-rate carries all the force dependence.
For an arbitrary value of $\gamma$, the mean-field solution does not agree with the exact result obtained for $N=2$.
This is expected since the mean-field approximation should work well only in the limit of large $N$.

The average velocity of the moving barrier is
\beq \label{eqn:vel} V=d(UN_0-Wq),\eeq
where the first term within the parenthesis is the contribution of the filaments  in contact  polymerizing with rate $U$
and the second term is the contribution from depolymerizing events of a single filament in contact. We have not found a way to solve in general the self-consistent equation satisfied by $q$, namely Eq.~\ref{self-consistent eq}, except near stalling conditions as explained in the next section. For this reason, we have calculated numerically $q$ from simulations, and
derived predictions from the mean-field theory assuming that $q$ is known.
For instance, using Eqs.~\ref{eqn:fullN0} and ~\ref{eqn:vel},
one obtains the average velocity.

\section{Results}

\subsection{Numerical validation of the mean-field approach}
We have tested the validity of the mean-field approach using numerical simulations. We used the classical Gillespie algorithm \cite{gillespie} incorporating the Mersenne Twister random number generator. Runs were executed for $N$ up to 5000. Up to 200 trial runs were used to derive averages and distributions.
We validated the simulation results by comparing them with the particular cases $N=1$ and for $N=2$ for which an exact solution is known (it is given in
\cite{david09} for $N=1$ and in the previous section for $N=2$).

By evaluating the parameter $q$ from the simulations, we obtained a very good agreement between the theoretical approach based on the use of mean-field and the simulations for the determination of the force velocity curve (shown in Fig.~\ref{fig:R5}, bottom) and
for the  number of filaments $N_0$ in contact with the barrier (shown in Fig.~\ref{fig:R5}, top). We find that the values of $N_i$ as determined by theory does not deviate from the simulation value by more than one.

\subsection{Condensation transition as function of the applied force}

\begin{figure}
\begin{center}
\includegraphics[scale=1.5]{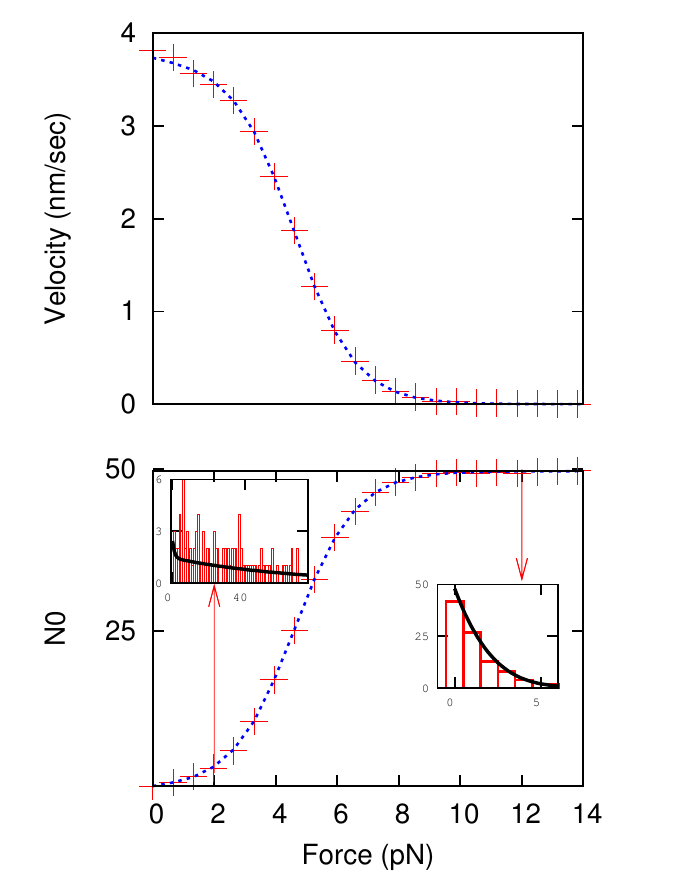}
\caption{Illustration of the condensation transition of actin filaments against a barrier.
\textit{Top}: Average barrier velocity vs. force. Symbols represent simulation results, the dotted line represents mean-field predictions based on Eq.~\ref{eqn:vel}.
\textit{Bottom}: Average number of filaments in contact with the barrier. Symbols represent simulation results, the dotted line represents mean-field predictions based on Eq.~\ref{eqn:fullN0}. For both plots, the parameters are $N=100$, $\gamma=1$ and $C=0.24\mu M$.
\textit{Inset, Left}: Density profile in the \emph{non-condensed} regime (bars) as function of the distance to the barrier, together with mean-field theory prediction (line) from Eq.~\ref{eqn:profile gen case}, for an applied force $F=2pN$, which is low with respect to the apparent stall force.
\textit{Inset, Right}: Density profile in the \emph{condensed} regime (bars) as function of the distance to barrier, together with mean-field theory prediction (line) from Eq.~\ref{eqn:profile gen case}, for an applied force $F=12$pN, which is close to the apparent stall force of $\approx12.5 pN$.}
\label{fig:R5}
\end{center}
\end{figure}

At low forces, the barrier velocity is close to its maximum value given by the free polymerization velocity.
In this case, only one or a small number of filaments are in contact, therefore $q \simeq 1$, which corresponds to
a \emph{non-condensed or single filament} regime.
The steady state density profile of the filaments is broad as shown in Fig.~\ref{fig:R5} (bottom, left inset) and the corresponding correlation length is large. With the parameters values corresponding to this figure, we have $l \simeq 151$nm.

Inversely, at high forces, the filaments accumulate at the barrier.
As a result $q \simeq 0$, the density profile is an exponential as shown in Fig.~\ref{fig:R5} (bottom, right inset) with a very short correlation length of the order of a monomer size. With the parameters values corresponding to this figure, we have $l \simeq 4.1$nm. Since in this case, the number of filaments  in contact, $N_0$ is a finite fraction of
$N$, we call this regime the \emph{condensed regime}.
In this high force regime (typically near the stall force $F=F_{stall}$), the following condition is obeyed:
$N U \ll U_0$. Since we also have $q \simeq 0$,
 Eq.~\ref{eqn:fullN0} simplifies to
\beq \label{eqn:N0} N_0=\left(1-\frac{W_0}{U_0}\right)N. \eeq
This equation can be used to predict the finite fraction of filaments  in contact  in the condensed regime. This condensed regime corresponds to the plateau
in the curve of $N_0$ vs. $F$ which is shown in Fig.~\ref{fig:R5} (top inset).
In the conditions of this figure, Eq.~\ref{eqn:N0} predicts a plateau
for $N_0 \simeq N/2=50$ which is indeed observed, and as expected the plateau
in $N_0$ (Fig.~\ref{fig:R5}, top) occurs at the same force
at which the velocity approaches zero (Fig.~\ref{fig:R5}, bottom).

\subsection{Theoretical stall force}
Let us first discuss here the theoretical expression of the stall force and then
in the next section the practical way this limit is approached.
The stall force is defined as the value of the force applied on the barrier for which
the velocity given by Eq.~\ref{eqn:vel} vanishes.
For $N=1$, the stall force is $F_{stall}^{(1)}=k_BT \ln ( U_0/W_0 )/d$. For $N=2$,
using the results obtained in Appendix A for $N_0$ and $q$, we find that the stall force $F_{stall}^{(2)}$, is exactly twice the stall force of a single filament,
$F_{stall}^{(1)}$, \beq \label{f2} F_{stall}^{(2)} =2 F_{stall}^{(1)}=2\frac{k_B T}{d} \ln\left(\frac{U_0}{W_0}\right). \eeq
In the general case of an arbitrary number of filaments $N$,
we expect that stall force $F_{stall}^{(N)}$ should be \cite{van-doorn,tanasethesis}:
\beq \label{eqn:stallN} F_{stall}^{(N)}=N \frac{k_B T}{d} \ln\frac{U_0}{W_0}. \eeq

This result can be derived from the following argument:
near stalling conditions, the average density of filaments at contact $N_0/N$, can be obtained from Eq.~\ref{eqn:N0} above. This average density of filaments can be used as an approximation of the probability to have one filament in contact when $N_0/N \ll 1$.
Since $q$ is the probability that there is a single filament in contact (in other words, there is
one filament among $N$ in contact and the remaining $N-1$ are free), it follows that
\beq \label{approx q} q={N \choose 1}\frac{N_0}{N} \left( 1 - \frac{N_0}{N} \right)^{N-1}, \eeq
which leads using Eq.~\ref{eqn:N0} to
\beq \label{eqn:qstall} q=N\left(1-\frac{W_0}{U_0}\right) \left(\frac{W_0}{U_0}\right)^{N-1} \simeq N_0\left(\frac{W_0}{U_0}\right)^{N-1}. \eeq
We call this the binomial form for q.
We note that Eq.~\ref{approx q} also means that
\beq \label{approx q Poisson} q \simeq N_0 \exp{(-N_0)}, \eeq
which corresponds to a Poisson statistics for the distribution of the number
of filaments at contact.
Now inserting the final expression for $q$ of Eq.~\ref{eqn:qstall} into the stalling condition, namely the vanishing of the velocity given by Eq.~\ref{eqn:vel}, one obtains the theoretical stall force given in Eq.~\ref{eqn:stallN}.

The theoretical expression of the stall force given by Eq.~\ref{eqn:stallN} has been also obtained in a recent study devoted to the stall force of a bundle of filaments \cite{kierfeld}. This study is based on the model introduced in Refs.~\cite{mog_ost,van-doorn} which the authors modified to include lateral interactions between the filaments of the bundle. Using a theoretical argument based on the identification of relevant polymerization cycles, the authors of Ref.~\cite{kierfeld} confirm the expression of the stall force obtained before in \cite{van-doorn}, which is also our Eq.~\ref{eqn:stallN}. More importantly, they show with this method that this expression has a universal character for models of this kind, hence in particular the independence of the stall force with respect to the load distribution factor $\gamma$. They also obtained force velocity curves for various values of the lateral interaction and staggering distance, which -as we have checked- agree with the numerical results obtained in this paper, when there is no lateral interaction and when the shifts are zero.

\begin{figure} 
\begin{center}
\includegraphics[scale=1.4]{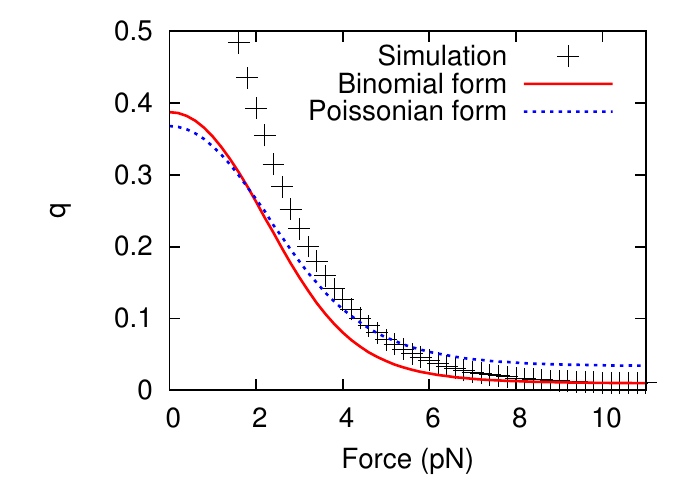}
\caption{Comparison between theoretical and numerical estimates for the parameter $q$, which represents the probability that there is a single filament in contact. Symbols represent simulation results, dotted line corresponds to Eq.~\ref{approx q Poisson} and continuous line corresponds to  Eq.~\ref{approx q} (both expressions are mean-field approximations valid in the high force regime). The parameters are $N=10$, $\gamma=1$ and $C=0.24\mu M$.}
\label{fig:Qf}
\end{center}
\end{figure}

\begin{figure}
\begin{center}
\includegraphics[scale=1.4]{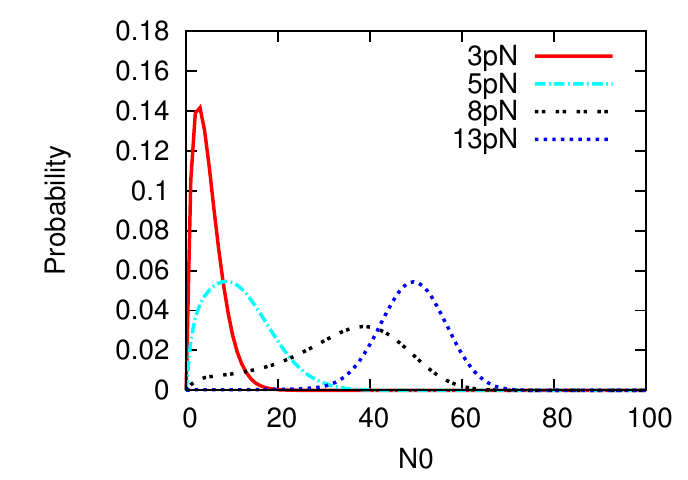}
\caption{Probability distributions of the number of filaments in contact with the barrier at various forces. The parameters are $N=100$,$\gamma=1$ and $C=0.24\mu M$.}
\label{fig:N0d100}
\end{center}
\end{figure}

In Fig.~\ref{fig:Qf}, the value of $q$ determined from the simulations is compared with theoretical expression given by Eq.~\ref{approx q} or Eq.~\ref{approx q Poisson} (both expressions give similar results). We note that the deviation between the simulation points and the theory increases as the force is lowered, this is due to the mean-field nature of the theory which becomes invalid when the force is small since then the fluctuations are large. For completeness, we also show in Fig.~\ref{fig:N0d100} the probability density function of the number of filaments at contact for various forces.

\subsection{The approach to stalling}
Let us now discuss more precisely how the velocity approaches zero.
We find that in our simulations, for $N$ larger than about 10, the velocity approaches zero at forces significantly lower than the stall force as shown in Fig.\ref{fig:R5} (bottom). We note that a similar effect has been obtained when analyzing the stall force of an
ensemble of interacting molecular motors \cite{campas}.
To quantify this effect, we therefore define an \emph{apparent} stall force, as the value of force where the velocity drops to less than a small fraction $\alpha=2.5\%$ of the value it has for zero force \cite{borisy}.
In the experimental situation, this bound could correspond for instance to the limit of resolution in the velocity measurement.

\begin{figure}
\begin{center}
\includegraphics[scale=0.35]{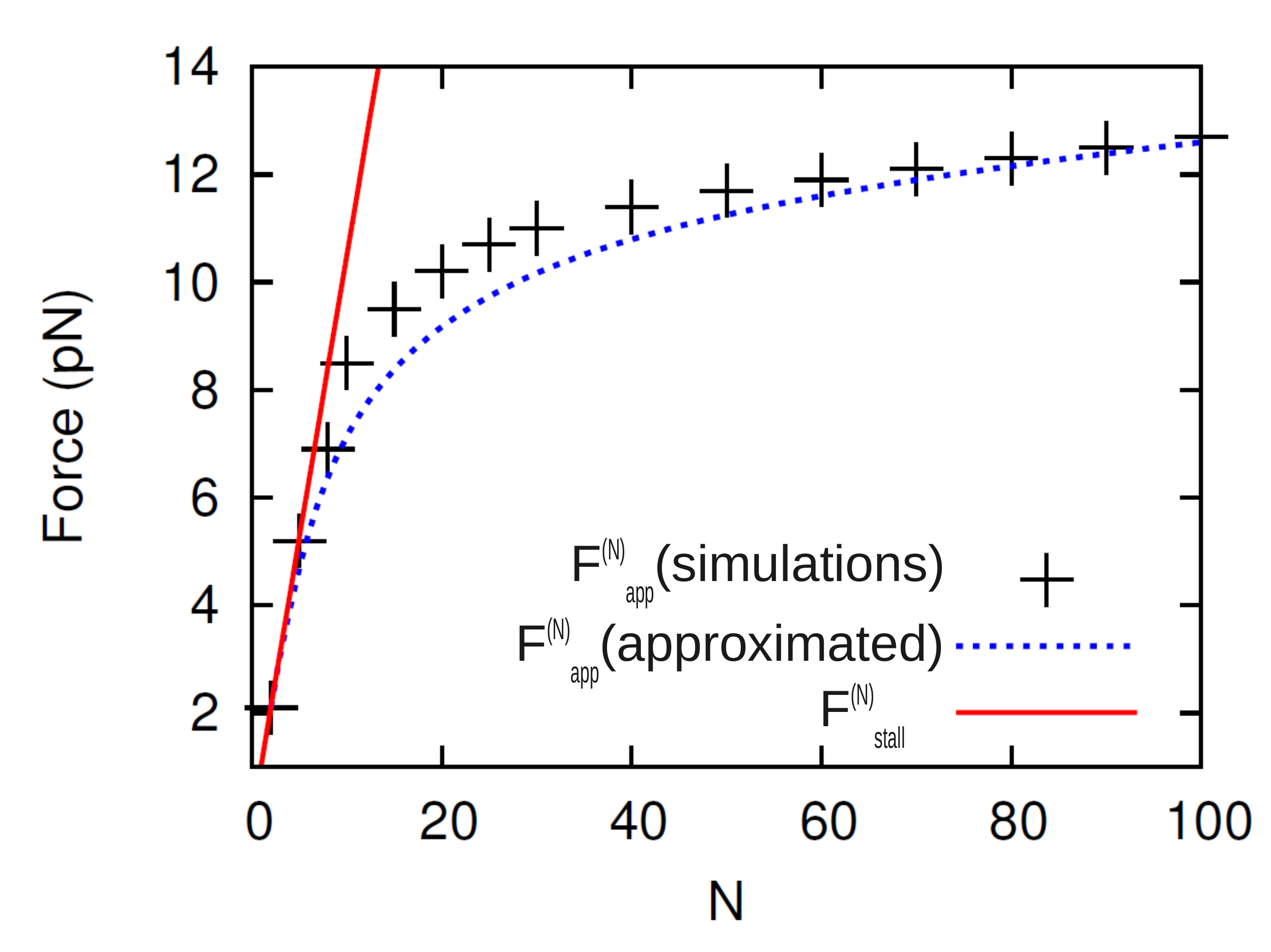}
\caption{Theoretical stall force $F^{(N)}_{stall}$ (straight line - calculated from Eq.~\ref{eqn:stallN}) and apparent stall force, both as computed from simulations $F^{(N)}_{simulations}$ (black symbols) and from the mean-field approximation given in Eq.~\ref{eqn:appstal} ($F^{(N)}_{approximated}$ -- dotted line) vs. number of filaments $N$. The parameters are $\gamma=1$ and $C=0.24\mu M$.}
\label{fig:Netsf}
\end{center}
\end{figure}

The value of the velocity at zero force corresponds to the maximum velocity. When $F=0$, there is no coupling between the filaments, which behave as independent random walkers. The probability to have more than one walker at the leading position is zero in the long time limit, which implies $q=1$. Therefore, $N_0=1$ and the velocity
at zero force equals the polymerization velocity of a single filament:
\beq
V(F=0)=d \left( U_0 - W_0 \right),
\eeq
which is mainly controlled by the monomer concentration.
Now using the expression of the velocity at an arbitrary force given by Eq.~\ref{eqn:vel},
the expression of $N_0$ given in Eq.~\ref{eqn:fullN0} and the parametrization of the rates of Eq.~\ref{eqn:rates} for the particular case $\gamma=1$, we find that
\beq \label{eqn:appstal} F_{app}^{(N)}=\frac{k_B T}{d} \ln{\frac{(1-\alpha)(U_0-W_0)N+\alpha U_0-(\alpha-q)W_0}{\alpha U_0-(\alpha-q)W_0}}. \eeq
Since $q \ll 1$ near stalling, we can write the following more
explicit expression
\beq
\label{eqn:appstallsimple} F_{app}^{(N)} \simeq \frac{k_B T}{d} \ln{ \left( 1+ \frac{N}{\alpha}-N \right) },
\eeq
In Fig.\ref{fig:Netsf}, we show the apparent stall force given by Eq.~\ref{eqn:appstal} as function of $N$ together with the theoretical stall force of Eq.~\ref{eqn:stallN}.

Let us show now that filament condensation at the barrier and the drop in velocity occur simultaneously.
Assuming for simplicity that $\gamma=1$, $N \gg 1$ and $q\simeq 0$ in the high force regime,
we can substitute Eq.~\ref{eqn:appstallsimple} into Eq.~\ref{eqn:fullN0} to obtain:
\beq N_0=(1-\alpha)\left(1-\frac{W_0}{U_0}\right)N. \eeq
From this we see that since $\alpha<<1$, the maximum number of filaments at the barrier is almost reached.
If $V_0$ is the initial velocity and $N_0^s$ is the finite fraction of filaments at the barrier at stall force, we have the equivalence of the following two conditions
\[V=\alpha V_0\Leftrightarrow N_0=(1-\alpha)N_0^s,\]
which shows that filament condensation occurs at the value of the apparent stall force, a point which is confirmed by simulations. Indeed, in the case of Fig.~\ref{fig:Netsf}
the apparent stall force is about 12.7 pN, and the
condensation visible in Fig.~\ref{fig:R5} also occurs close to $12$ pN.

\begin{figure}
\begin{center}
\includegraphics[scale=1.4]{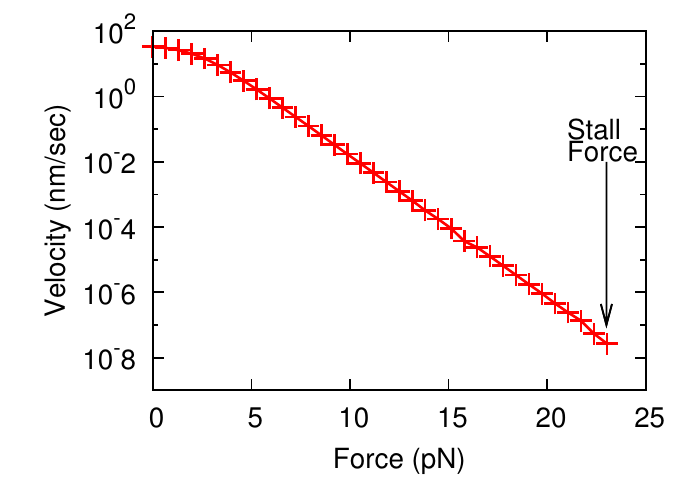}
\caption{Average barrier velocity in logarithmic scale as function of the force in linear scale. Note that the velocity
decreases to near zero exponentially when approaching the theoretical stall force, which is shown with the arrow.
For this value of the force, the numerical velocity is not strictly zero but it is close to the uncertainty intrinsic to the simulation, which is here of the order of $10^{-8}$nm/s. The parameters are $N=10$, $\gamma$=1 and $C=1.2\mu M$.}
\label{fig:sflog}
\end{center}
\end{figure}

Close to stall force it is also possible to derive an analytic expression
for the force-velocity relation by substituting into
Eq.~\ref{eqn:vel} the expressions of $q$, given by Eq~\ref{approx q} and Eq.~\ref{approx q Poisson}. Assuming for simplicity $\gamma=1$, and using Eq.~\ref{eqn:N0}, we obtain with the binomial form:
\beq V=N\left(1-\frac{W_0}{U_0}\right)\left[U_0e^{-f}-W_0\left(\frac{W_0}{U_0}\right)^{N-1}\right],\eeq
and with the Poissonian form:
\beq V=N\left(1-\frac{W_0}{U_0}\right)\left(U_0e^{-f}-W_0e^{-N(1-W_0/U_0)}\right).\eeq
When these expressions are expanded close to stall force, one obtains in both cases:
\beq \delta V=N\left(1-\frac{W_0}{U_0}\right)U_0e^{-f}\delta f.\eeq
This indicates an exponential dependence of the velocity close to stalling, which is indeed present in the simulations as shown in Fig.\ref{fig:sflog}.

To summarize, we have shown in this section that the apparent stall force does not scale linearly with $N$ as the theoretical stall force but rather as $\ln{(N)}$. The apparent stall force is the quantity of experimental interest, it is also near the apparent stall force that the condensation transition discussed in a previous section occurs (nothing special of that sort occurs near the theoretical stall force).

\subsection{Related experimental work in connection with the model}

In this section we discuss related experimental work. Although a precise comparison with the present model is not attempted,
we hope that the discussion could be useful in identifying some relevant questions in this field.
The force generation by parallel actin filaments growing out of an acrosome bundle
has been measured in Ref.~\cite{footer}. The observation of a plateau in force measurements by optical tweezers is a good indication of the stalling regime, but the measured stall force is very small, comparable with that of a single filament, although many filaments are present (about a dozen). These results thus stand at odds with the theoretical predictions for the stall force obtained in Refs.~\cite{van-doorn,kierfeld} (and in the present paper).
In the present paper, we have emphasized the fact that the approach to stalling is slow, which can lead to an underestimation of the true stall force. The resolution of the optical tweezers leads to a limit in the detection of small velocities, which corresponds roughly to the criterion for the apparent stall force used in the previous section. However, with a dozen filaments, the apparent stall force should be significantly larger than that of a single filament. Another difficulty is that there is no indication in this experiment of the two regimes of low and large forces discussed in this paper. At this point, it may be important to say that the results of this experiment have not been reproduced, in fact in a new experiment
discussed below, where the force generated by filaments growing from two magnetic beads outwards, very different results have been obtained \cite{experiment}.
In view of all this, we think that the reason for these discrepancies may be foundin effects which are not accounted for (such as buckling or filament cross-linking) or they maybe attributed more simply to the fact that the two experiments have been done in very different biochemical conditions. Indeed, the authors of Ref.~\cite{footer} have used profilin to suppress spontaneous nucleation of actin filaments, while profilin was absent in \cite{experiment}. The use of profilin in Ref.~\cite{footer} introduced complications since profilin also modifies the thermodynamics of the system by binding to actin monomers, and possibly interferes with ATP hydrolysis during polymerization.

The mechanical response of an actin networks confined between two rigid flat surfaces has been probed using a surface force apparatus (SFA) in Ref.~\cite{israel}, and using an atomic force microscope (AFM) in \cite{fletcher}. Both experiments reported a load history dependent mechanical response, which presumably reflects a complex interplay between buckling and polymerization forces.
This complex interplay makes it difficult to isolate the true contribution of polymerization forces.
More recently, C. Brangbour et al. devised a new experimental setup in which actin is nucleated from magnetic beads which are covered by gelsolin \cite{experiment}. A magnetic field is used to counteract the polymerization force, which allows to measure the force-velocity curves. As mentioned above, the results of Ref~\cite{footer} for the stall force of a single filament are not confirmed: on the contrary, the stall force which is obtained is of the order of 40 pN, which corresponds according to Eq.~\ref{eqn:stallN} to about 25 active filaments. The general shape of these force-velocity curves is similar to the ones obtained in this work, but some deviations are present at low and high forces. These discrepancies suggest that our model may be too simple to fully explain this experiment, and that other aspects may be important. First, it would be necessary to go beyond the parallel organization of the filaments in order to better model the experimental geometry of Ref.~\cite{experiment}. Secondly, it is probably important to account in the model for the possibility of nucleating new filaments from existing ones \cite{blanchoin10}. Thirdly, buckling forces could play an important role in the experiment. Some of these effects have been included in previous numerical simulations of branched actin networks \cite{carlsson,lee_liu09}, but they are typically difficult to study with analytical models of the kind presented here.

\section{Conclusion}
In this paper, we have provided a new theoretical framework to describe the dynamics of an ensemble of $N$ parallel filaments with no lateral interactions, which are exerting a force against a movable barrier.
The special cases $N=1$ and $N=2$ can be solved exactly, unlike the general case for arbitrary $N$, for which we have constructed a mean-field approach.
We identify two regimes: a non-condensed regime at low force in which filaments are spread out spatially, and a condensed regime at high force in which filaments accumulate near the barrier. The transition occurs near the apparent stall force where the velocity approaches zero.
We find that for large $N$ this regime where velocity approaches zero occurs at forces significantly lower than the theoretical stall force, given by $N$ times the stall force of one filament. In fact, the apparent stall force does not scale linearly with $N$ as the theoretical stall force does; instead it scales logarithmically.

On the theory side, several extensions of our work are worth investigating.
For instance, bundles can be formed experimentally by growing filaments in the presence of specific proteins which cross-link the filaments. To describe such a situation, it would be necessary to include lateral interactions.
Another direction would be to explore the role of load sharing, as done in \cite{borisy} for instance. Although the dynamics will be different, we still expect a condensation transition to be present in this case.

In the end, our model offers a very simplified view of the problem of force generation by actin filaments, but precisely for this reason we hope that it can be a useful starting point for more refined studies.

\ack
The authors would like to thank A.B. Kolomeisky, P. Sens, R. Pandinhateeri for stimulating discussions, and J. Baudry for a careful reading of the manuscript. K. Tsekouras would also like to thank J. Elgeti for his
help with computational issues. This work has been supported by the ANR (french
national research agency) under contract ANR-09-PIRI-0001-02.

\begin{appendix}
\section*{Appendix: Exact solution of the master equation for case ($N=2$)}
\setcounter{section}{1}
For this problem, we can derive the following master equation satisfied by $p(n,t)$ the probability
that there is a gap of $n$ monomers between the two filaments. For $n>2$,
\beq \pd{p(n,t)}{t} = (U+W_0)[p(n-1,t)-p(n,t)]+(W+U_0)[p(n+1,t)-p(n,t)], \label{sol N=2} \eeq
and otherwise,
\beq \pd{p(1,t)}{t} = 2(U+W_0)p(0,t)-(U+W_0)p(1,t)+(W+U_0)[p(2,t)-p(1,t)], \label{sol N=2b} \eeq
\beq \pd{p(0,t)}{t} = (W+U_0)p(1,t)-2(U+W_0)p(0,t). \label{sol N=2bb} \eeq
Solving Eq.~\ref{sol N=2} at steady state results in the recursion
\[(W+U_0)p(n+1)-(U+W+U_0+W_0)p(n)+(U+W_0)p(n-1)=0,\]
which yields two solutions namely 1 and $b=\frac{U+W_0}{W+U_0}$.
This means that for $n\ge 2$, $p(n)=p(2) b^{n-2}$.

Using the normalization condition:
$p(0)+p(1)+\sum_{n\ge 2}p(2)b^{n-2}=1$, we obtain
$p(2)=(1-b) (1-p(0)-p(1))$.
Solving Eq.~\ref{sol N=2bb} at steady state results in:
\[ p(1)=2p(0)\frac{U+W_0}{W+U_0}=2p(0)b. \]
Substituting this expression into Eq.~\ref{sol N=2b} at steady state yields:
\[p(2)=2p(0)b^2.\]
Equating the two expressions for $p(2)$ and using the expression for $p(1)$ in terms of $p(2)$, it follows that
\[p(0)=\frac{1-b}{1+b}.\]
The probability of having a gap of zero monomers is the probability of having both filaments at the barrier; it thus obeys
$p(0)+q=1$, since $q$ is the probability of having only one filament at the barrier. In the end, we find
\beq q=\frac{2(U+W_0)}{U+W+U_0+W_0}.\eeq
The average number of filaments at the barrier is the sum of $q$ plus twice $1-q$,
since $1-q$ is the probability of having both filaments at the barrier. Therefore:
\[N_0=q+2(1-q)=2-q,\]
and if we substitute $q$ from above we find
\beq N_0=\frac{2(W+U_0)}{U+W+U_0+W_0}.\eeq
\end{appendix}

\clearpage

\clearpage

\clearpage

\clearpage

\clearpage

\clearpage

\clearpage

\section{References}
\bibliography{lkj}{}

\begin{thebibliography}{10}

\bibitem{pollard09}
{Pollard T.D.} and {J.A. Cooper}.
\newblock Actin, a central player in cell shape and movement.
\newblock {\em Science}, 326, 2009.

\bibitem{kovar_pollard}
{Kovar R.D.} and {T.D. Pollard}.
\newblock Insertional assembly of actin filament barbed ends in association
  with formins produces piconewton forces.
\newblock {\em Proc. Natl. Acad. Sci. USA.}, 41:14725--14730, 2004.

\bibitem{mitchison}
{Kueh H.Y.} and {T.J. Mitchison}.
\newblock Structural plasticity in actin and tubulin polymer dynamics.
\newblock {\em Science}, 325, 2009.

\bibitem{kolomeisky05}
{Stukalin E.B.} and {A.B. Kolomeisky}.
\newblock Polymerization dynamics of double-stranded biopolymers: Chemical
  kinetic approach.
\newblock {\em J. Chem. Phys.}, 122:104903, 2005.

\bibitem{kolomeisky06}
{Stukalin E.B.} and {A.B. Kolomeisky}.
\newblock {ATP} hydrolysis stimulates large length fluctuations in single actin
  filaments.
\newblock {\em Biophys. J.}, 90(8):2673--2685, 2006.

\bibitem{david09}
{Ranjith P., D. Lacoste, K. Mallick} and {J.-F. Joanny}.
\newblock Nonequilibrium self-assembly of a filament coupled to {ATP/GTP}
  hydrolysis.
\newblock {\em Biophys. J.}, 96:2146--2159, 2009.

\bibitem{david10}
{Ranjith P., D. Lacoste, K. Mallick} and {J.-F. Joanny}.
\newblock Role of {ATP}-hydrolysis in the dynamics of a single actin filament.
\newblock {\em Biophys. J.}, 98:1418--1427, 2010.

\bibitem{vavylonis05}
{Vavylonis D., O. Yang} and {B. O'Shaughnessy}.
\newblock {Actin polymerization kinetics, cap structure, and fluctuations}.
\newblock {\em Proc. Natl. Acad. Sci. USA.}, 102(24):8543--8548, 2005.

\bibitem{antal-ideal}
{Antal T., P.L. Krapivsky, S. Redner, M. Mailman} and {B. Chakraborty}.
\newblock Dynamics of an idealized model of microtubule growth and catastrophe.
\newblock {\em Phys. Rev. E.}, 76(4):041907, 2007.

\bibitem{antal-instab}
{Antal T., P.L. Krapivsky} and {S. Redner}.
\newblock Dynamics of microtubule instabilities.
\newblock {\em J. Stat. Mech.}, page L05004, 2007.

\bibitem{lipowsky}
{Li X., R. Lipowsky} and {J. Kierfeld}.
\newblock Coupling of actin hydrolysis and polymerization: Reduced description
  with two nucleotide states.
\newblock {\em Eur. Phys. Lett.}, 89:38010, 2010.

\bibitem{altigan_sun}
D.~Wirtz Altigan~E. and S.X. Sun.
\newblock Mechanics and dynamics of actin-driven thin membrane protrusions.
\newblock {\em Biophys. J.}, 90:65--76, 2006.

\bibitem{hill_kir}
{Hill T.L} and {M.W. Kirschner}.
\newblock Subunit treadmilling of microtubules or actin in the presence of
  cellular barriers: Possible conversion of chemical free energy into
  mechanical work.
\newblock {\em Proc. Natl. Acad. Sci. USA.}, 79:490--494, 1981.

\bibitem{mog_ost}
{Mogilner A.} and {G. Oster}.
\newblock The polymerization ratchet model explains the force-velocity relation
  for growing microtubules.
\newblock {\em Eur. Biophys. J.}, 28:235--242, 1999.

\bibitem{van-doorn}
{Sander van Doorn G., C. Tanase, B.M. Mulder} and {M. Dogterom}.
\newblock On the stall force for growing microtubules.
\newblock {\em Eur. Biophys. J.}, 20:2--6, 2000.

\bibitem{tanasethesis}
{Tanase C.}
\newblock {\em Physical modeling of microtubule force generation and
  self-organization}.
\newblock PhD Thesis, Wageningen University, 2004.

\bibitem{note}
{Vavylonis D.} and {B. O'Shaughnessy}. ``Brownian ratchet model of force
  generation and kinetics of actin filament bundles." Presented at the 50th
  Annual Meeting of the Biophysical Society, Feb. 18-22 2006, Salt Lake City,
  USA.

\bibitem{carlsson}
{Carlsson A.E.}
\newblock Growth velocities of branched actin networks.
\newblock {\em Biophys. J.}, 84:2907--2918, 2003.

\bibitem{peskin_oster93}
{Peskin C.S., G.M. Odell} and {G.F. Oster}.
\newblock Cellular motions and thermal fluctuations: the brownian ratchet.
\newblock {\em Biophys. J.}, 65(1):316--324, 1993.

\bibitem{kierfeld}
{Krawczyk J.} and {J. Kierfeld}.
\newblock Stall force of polymerizing microtubules and filament bundles.
\newblock {\em Europhys. Lett.}, 93:28006, 2011.

\bibitem{footer}
{Footer M.J., J.W.J. Kerssemakers, J.A. Theriot} and {M. Dogterom}.
\newblock Direct measurement of force generation by actin filament
  polymerization using an optical trap.
\newblock {\em Proc. Natl. Acad. Sci. USA.}, 104(7):2181--2186, 2006.

\bibitem{experiment}
{Brangbour C., O. du Roure, E. Helfer, D. D{\'e}moulin, A. Mazurier, M.
  Fermigier, M.-F. Carlier, J. Bibette} and {J. Baudry}.
\newblock Force velocity measurements of a few growing actin filaments.
\newblock {\em PLoS Biol}, 9 (4):e1000613, 2011.

\bibitem{hill-book}
{Hill T.L.}
\newblock {\em Linear Aggregation Theory in Cell biology}.
\newblock Springer, Berlin, Germany, 1987.

\bibitem{bell}
{Bell G.I}.
\newblock Models for the specific adhesion of cells to cells.
\newblock {\em Science}, 200, 1978.

\bibitem{walcott}
{Walcott S.}
\newblock The load dependence of rate constants.
\newblock {\em J. Chem. Phys.}, 128:215101, 2008.

\bibitem{borisy}
{Schaus T.E.} and {G.G. Borisy}.
\newblock Performance of a population of independent filaments in lamellipodial
  protrusion.
\newblock {\em Biophys. J.}, 95:1393--1411, 2008.

\bibitem{gillespie}
{Gillespie D.T.}
\newblock A general method for numerically simulating the stochastic time
  evolution of coupled chemical reactions.
\newblock {\em J. Comp. Phys.}, 22:403--434, 1976.

\bibitem{campas}
{Camp{\'a}s O., Y. Kafri K. B. Zeldovich J. Casademunt} and {J.-F. Joanny}.
\newblock Collective dynamics of interacting molecular motors.
\newblock {\em Phys. Rev. Lett.}, 97:038101, 2006.

\bibitem{israel}
{Greene G.W., T.H. Anderson, H. Zen, B. Zappone} and {J.N. Israelachvili}.
\newblock Force amplification response of actin filaments under confined
  compression.
\newblock {\em Proc. Natl. Acad. Sci. USA.}, 106:445--449, 2008.

\bibitem{fletcher}
{Fletcher D.}
\newblock Loading history determines velocity of actin network growth.
\newblock {\em Nature Cell Biol.}, 7:1219--1223, 2005.

\bibitem{blanchoin10}
{Achard V., J.-L. Martiel, A. Michelot, C. Guerin, A.-C. Reymann, L. Blanchoin}
  and {R. Boujemaa-Paterski}.
\newblock A primer-based mechanism underlies branched actin filament network
  formation and motility.
\newblock {\em Curr. Biol.}, 20:423--428, 2010.

\bibitem{lee_liu09}
{Lee K.-C.} and {A.J. Liu}.
\newblock Force-velocity relation for actin-polumerization-driven motility from
  brownian dynamics simulations.
\newblock {\em Biophys. J.}, 97:1295--1304, 2009.

\end{thebibliography}
\bibliographystyle{unsrt}

\clearpage

\begin{table*}
\caption{Parameters characterizing an actin filament barbed end. $W_0$ is the free filament depolymerization rate, $k_0$ is the rate constant entering the free filament polymerization rate $U_0=k_0 C$, where $C$ is the concentration of free monomers, $d$ is the monomer size and $C_c$ is the critical concentration.}
\label{key}
\centering
\begin{tabular}{ c  c  c  c }
\br
$W_0 (s^{-1})$  & $k_0 (\mu M^{-1}s^{-1})$ & d(nm) & $C_c (\mu M)$\\
\mr
1.4 & 11.6 & 2.7 & 0.141\\
\br
\end{tabular}
\end{table*}

\end{document}